\documentclass[aps,nofootinbib,preprint,showpacs,showkeys]{revtex4}
\usepackage{graphicx,color,amsmath}
\begin{document}

\title{Bottomed Analog of $Z^+(4433)$}
\author{Kingman Cheung$^{a,b}$, Wai-Yee Keung$^{c}$ and Tzu-Chiang Yuan$^{b}$}
\affiliation{
$^a$ Department of Physics, National Tsing Hua University, Hsinchu 300, Taiwan
 \\
$^b$ National Center for Theoretical Sciences, Hsinchu 300, Taiwan \\
$^c$ Department of Physics, University of Illinois, Chicago IL 60607-7059, USA
}
\date{\today}

\begin{abstract}
The newly observed $Z^+ (4433)$ resonance by BELLE 
is believed to be a tetraquark bound state made up of $(cu)(\bar c \bar d)$.
We propose the bottomed analog of this bound state, namely, by 
replacing one of the charm quarks by a bottom quark, 
thus forming $Z_{bc}^{0, \pm, \pm\pm}$.  One of the $Z_{bc}$ is doubly
charged.  The predicted mass of $Z_{bc}$ is around 7.6 GeV.
This doubly-charged bound state can be detected by its decay into 
$B_c^\pm  \pi^\pm$.
Similarly, we can also replace both charm quark and antiquark of
the $Z^+ (4433)$ by bottom quark and antiquark, respectively,  thus
forming $Z_{bb}$ the bottomonium analog of $Z^+ (4433)$. The predicted mass
of $Z_{bb}$ is about 10.7 GeV.
\end{abstract}
\maketitle

\section{Introduction}

A recent observation of a new resonance state, denoted by
$Z^+ (4433)$, by the BELLE Collaboration \cite{belle} via the decay channel
\begin{equation}
  Z^+ (4433) \to \psi'\; \pi^+ 
\end{equation}
may be interpreted as
a tetraquark bound state made up of diquark-antidiquark
$(c u)(\bar c \bar d)$  \cite{Maiani}.
However, there are alternative views that the observed peak is instead  a
rescattering effect of the $D^*(2010) \overline{D}_1 (2420)$ molecule
\cite{rosner,Meng}, because of the closeness to the
$D^*(2010)  \overline{D}_1 (2420)$ production threshold.  
It remains unclear at the moment whether it is a genuine tetraquark or
merely a rescattering effect.

If the observed peak is a genuine tetraquark, it is the first 
time that a charged tetraquark bound state is observed. 
Unlike the interpretations of other possible tetraquark bound-state
candidates \cite{genuine}, 
this charged tetraquark bound state is unambiguously 
interpreted as a 4-quark bound state, because it cannot arise from
higher Fock states of charmonium. 
Maiani {\it et al.} \cite{Maiani} interpreted the new resonance 
as an orbitally excited
state of $X(3872)$ and $X(3876)$. One of the reasons is that
the mass difference between $Z^+ (4433)$ and $X(3872)$ is close to
the mass difference between $\psi$ and $\psi'$.
Another important property of this new state is its narrow width \cite{belle}
\[
\Gamma = 44 \; ^{+17}_{-13} \;({\rm stat}) \; ^{+30}_{-11}
    \; ({\rm syst}) \;\; {\rm MeV} \;,
\] 
which suggests that it is not likely a loosely bound molecule of
$D^* (2010)  \overline{D}_1 (2420)$, but perhaps a genuine tetraquark state 
\cite{Maiani}.

In this note, we suggest to look for the bottomed analog of $Z^+(4433)$ 
by replacing one of the charm quarks by a bottom quark.
We denote these bottomed tetraquarks by $Z_{bc}$.  It can have a 
few charged states: $Z_{bc}^{0, \pm, \pm\pm}$. This suggestion is valid in 
the constituent quark model. The most 
interesting result is the existence of the
doubly-charged $Z_{bc}^{\pm\pm}$.  We can even replace both the
charm quark and antiquark of $Z^+ (4433)$ by the bottom quark and antiquark,
thus forming $Z_{bb}$ the bottomonium analog of $Z^+ (4433)$. 
We show in Table I the quark contents of these tetraquark bound states
along with their charges and isospins.  

\begin{table}[th!]
\caption{\small Quark contents of $Z (4433)$ and the bottomed analog
$Z_{bc}$ and the bottomonium analog $Z_{bb}$, $(Q= I_3 + Y)$.
The charge conjugated states of $Z_{bc}^{0,-,--}$ are not shown, but
can be easily obtained.
Normalization factor $1/\sqrt{2}$ is omitted for states of $I_3=0$.
}
\begin{ruledtabular}
\begin{tabular}{l|r|l|l|l}
$I$   &
$I_3$ &
$Z(4433),\ Y=0$   &  $Z_{bc},\ Y=-1$  & $Z_{bb},\ Y=0$ \\
\hline
$0$ & $0$ &
$Z_{cc}^0:\ (cu)(\bar c \bar u)+(cd)(\bar c \bar d)$ & 
$Z_{bc}^-:\ (bu)(\bar c \bar u)+(bd)(\bar c \bar d)$ &
$Z_{bb}^0:\ (bu)(\bar b \bar u)+(bd)(\bar b \bar d)$ 
\\ \hline
& $0$&
$Z_{cc}^0:\ (cu)(\bar c \bar u)-(cd)(\bar c \bar d)$ & 
$Z_{bc}^-:\ (bu)(\bar c \bar u)-(bd)(\bar c \bar d)$ &
$Z_{bb}^0:\ (bu)(\bar b \bar u)-(bd)(\bar b \bar d)$ \\
1 &$+1$&
$Z_{cc}^+:\;(cu)(\bar c \bar d)$ & $Z_{bc}^0:\;(bu)(\bar c \bar d)$ &
    $Z_{bb}^+:\;(bu)(\bar b \bar d)$\\
& $-1$&
$Z_{cc}^-:\;(cd)(\bar c \bar u)$ & $Z_{bc}^{--}:\;(bd)(\bar c \bar u)$ &
   $Z_{bb}^-:\;(bd)(\bar b \bar u)$ 
\end{tabular}
\end{ruledtabular}
\end{table}

\section{Mass Estimation}
To get a crude estimate of the mass of the $Z_{bc}$ states, we 
naively replace the charm quark mass by a bottom quark mass.  
We thus expect that the mass difference between
$B_c$ and $\psi$ mesons of the same radial-orbital quantum numbers are 
nearly the same.
The masses of $B_c$ mesons cited below are taken from Ref. \cite{Eichten}.
The experimental value \cite{pdg} of the mass of the lowest $B_c$ meson 
is very close to the prediction in Ref. \cite{Eichten}
based on Buchm\"{u}ller-Tye potential \cite{BT}.
Experimentally, the following two mass differences
\begin{eqnarray}
\label{3213}
&& M \left[ B_c (2\, ^3 S_1) \right] - 
M \left[ \psi (2 \,^3 S_1) \right] = 6899 - 3686 \; {\rm MeV}  
= 3213 \; {\rm MeV} 
\end{eqnarray}
and
\begin{eqnarray}
\label{3240}
M \left[ B_c (1\, ^3 S_1) \right] - 
M \left[ \psi (1\, ^3 S_1) \right] & = & 6337 - 3097 \; {\rm MeV} = 
3240 \; {\rm MeV} 
\end{eqnarray}
are indeed quite proximate. We thus expect 
\begin{equation}
\label{3210}
M \left[ Z_{bc}\right] - 
M \left[ Z_{cc}\right] \simeq  3200 \; {\rm MeV}
\end{equation}
with an uncertainty of the order of $50-100$ MeV.
Therefore for $I=1$, we predict the mass of the bottomed analog $Z_{bc}$ to be
\begin{equation}
\label{7640}
M \left[ Z_{bc}\right]  = 7630 \pm 100 \;{\rm MeV}\;.
\end{equation}
Using the same approach the mass of the bottomonium analog $Z_{bb}$ meson is
predicted to be
\begin{eqnarray}
M \left[ Z_{bb}\right]  &=& M \left[ Z_{cc}\right]  +
\left( M \left[ \Upsilon( 2\, ^3S_1) \right ] - 
       M \left[ \psi( 2\, ^3S_1) \right ]    \right )  \nonumber \\
 &=& M \left[ Z_{cc}\right]  + 6300 \; {\rm MeV} 
 = 10730 \pm 100 \; {\rm MeV} \; .
\end{eqnarray}

\section{Discussions}

In Ref. \cite{rosner}, it was suggested that the observed peak of 
$Z^+ (4433)$ is a rescattering effect of $D^*(2010) \overline{D}_1(2420)$
because of the closeness of the $D^*(2010) \overline{D}_1(2420)$ threshold.
Analogously, the estimated mass of $Z_{bc}$ in Eq. (\ref{7640})
is also close to the threshold of
$B (5279) D_1(2420)$ or $B^* (5325) D_0^*(2400) \sim 7700 $ MeV, within the uncertainty
of our estimation.  

Current experiments that can search for the proposed $Z_{bc}$ or
$Z_{bb}$ bound states are the CDF and D\O\, because of their higher masses.  
The discovery channel would be
\begin{equation}
  Z_{bc}^{++} \to B_c^{+} (2\, ^3S_1) \; \pi^+ \;, 
\end{equation}
where the $B_c^+\;(2 \, ^3 S_1)$ is in turn detected by its decay 
into $B_c^+\;(1\, ^3S_1)+ \pi\pi$ \cite{Eichten}.
We have no prediction for the production rate, because of the complicated
fragmentation and recombination effects in forming the tetraquark state.

Similarly, the bottomonium analog can be searched for via the channel
\begin{equation}
 Z^+_{bb} \to \Upsilon(2S) \; \pi^+ \,.
\end{equation}

In summary, we have proposed the bottom and bottomonium analogs of
the
newly observed $Z^+ (4433)$ meson at BELLE. 
The most interesting finding is the existence of doubly-charged
$Z_{bc}^{++}$ state, which can be searched in the channel $B_c^+
\pi^+$.
Just like the $Z^+ (4433)$, these analogs with isospin $I=1$
 are expected to be in the
$2S$ states having $J^{PC}=1^{+-}$. The $1S$ states and states with
other $J^{PC}$ assignments are also expected.
Extension of the present analysis to the full $SU(3)$ symmetry
including strangeness is straightforward and will not be presented
here.

This research was supported in parts by the NSC
under Grant No.\ NSC 96-2628-M-007-002-MY3, by the NCTS,
and by U.S. DOE under Grant No. DE-FG02-84ER40173.

\end{document}